\documentclass[conference]{IEEEtran}

\usepackage[numbers,sort&compress]{natbib} % Replaces 'cite' and handles 'citep'
\usepackage{amsmath,amssymb,amsfonts}
\usepackage{amsthm}
\usepackage{mathrsfs}
\usepackage{graphicx}
\usepackage{textcomp}
\usepackage{xcolor}
\usepackage{times}
\usepackage{algorithm}
\usepackage{algorithmic} % Only need this once

\usepackage{booktabs}
\usepackage{array}
\usepackage{multirow}
\usepackage{tabularx}
\usepackage{float}
\usepackage{subcaption}
\usepackage{caption}
\usepackage{adjustbox}

\usepackage{tikz}
\usetikzlibrary{arrows.meta, arrows, positioning, shapes.geometric, fit, svg.path}

\IfFileExists{pgfplots.sty}{
    \usepackage{pgfplots}
    \pgfplotsset{compat=1.18}
}{}

\usepackage{url}
\usepackage[colorlinks=true, linkcolor=black, citecolor=black, urlcolor=blue]{hyperref}
\usepackage{comment}
\usepackage{blindtext}
\usepackage{lipsum}
\usepackage{mwe}

\IEEEoverridecommandlockouts

\makeatletter
\newcommand{\fixedcaption}[2]{%
    \def\@captype{#1}%
    \caption{#2}%
}
\makeatother

\def\BibTeX{{\rm B\kern-.05em{\sc i\kern-.025em b}\kern-.08em
    T\kern-.1667em\lower.7ex\hbox{E}\kern-.125emX}}

\theoremstyle{plain}

\tikzset{
    startstop/.style={rectangle, rounded corners, minimum width=7cm, minimum height=1cm, text centered, draw=black},
    process/.style={rectangle, minimum width=4cm, minimum height=1.8cm, text centered, draw=black},
    arrow/.style={thick,->,>=stealth}
}

\usepackage{scalerel}
\definecolor{orcidlogocol}{HTML}{A6CE39}
\tikzset{
  orcidlogo/.pic={
    \fill[orcidlogocol] svg{M256,128c0,70.7-57.3,128-128,128C57.3,256,0,198.7,0,128C0,57.3,57.3,0,128,0C198.7,0,256,57.3,256,128z};
    \fill[white] svg{M86.3,186.2H70.9V79.1h15.4v48.4V186.2z}
                 svg{M108.9,79.1h41.6c39.6,0,57,28.3,57,53.6c0,27.5-21.5,53.6-56.8,53.6h-41.8V79.1z}
                 svg{M124.3,172.4h24.5c34.9,0,42.9-26.5,42.9-39.7c0-21.5-13.7-39.7-43.7-39.7h-23.7V172.4z}
                 svg{M88.7,56.8c0,5.5-4.5,10.1-10.1,10.1c-5.6,0-10.1-4.6-10.1-10.1c0-5.6,4.5-10.1,10.1-10.1C84.2,46.7,88.7,51.3,88.7,56.8z};
  }
}
\newcommand\orcidicon[1]{\href{https://orcid.org/#1}{\mbox{\scalerel*{
\begin{tikzpicture}[yscale=-1,transform shape]
\pic{orcidlogo};
\end{tikzpicture}
}{|}}}}

\begin{document}

\title{\LARGE Can Cross-Layer Design Bridge Security and Efficiency? A Robust Authentication Framework for Healthcare Information Exchange Systems}

% \author{\authorblockN{Leave Author List blank for your IMS2013 Summary (initial) submission.\\ IMS2013 will be rigorously enforcing the new double-blind reviewing requirements.}
% \authorblockA{\authorrefmark{1}Leave Affiliation List blank for your Summary (initial) submission}}

\author{
    \IEEEauthorblockN{
        Khalid M. Ezzat\IEEEauthorrefmark{1}\orcidicon{0009-0005-6215-8596}, 
        Muhammad El-Saba\IEEEauthorrefmark{2}, and 
        Mahmoud A. Shawky\IEEEauthorrefmark{3}\IEEEauthorrefmark{4}\orcidicon{0000-0003-3393-8460}
    }

    \IEEEauthorblockA{
        \IEEEauthorrefmark{1}Department of Electronics and Communications, Air Defense College, Military Academy, Cairo, Egypt\\
        Email: 2100021@eng.asu.edu.eg
    }

    \IEEEauthorblockA{
        \IEEEauthorrefmark{2}Department of Electronics and Communications, Faculty of Engineering, Ain Shams University, Cairo, Egypt\\
        Email: melsaba@eng.asu.edu.eg
    }

    \IEEEauthorblockA{
        \IEEEauthorrefmark{3}School of Computer Science and Electronic Engineering, University of Essex, Colchester, UK\\
        Email: mahmoud.a.shawky-ahmed@essex.ac.uk
    }

    \IEEEauthorblockA{
        \IEEEauthorrefmark{4}The Egyptian Technical Research and Development Center, Cairo, Egypt
    }
    
}

\maketitle

\begin{abstract}
As healthcare systems become increasingly interconnected, ensuring secure and continuous device authentication in health information exchange (HIE) networks is critical to safeguarding patient data and clinical operations. In this context, this paper proposes a novel cross-layer authentication scheme for HIE networks that integrates cryptographic mechanisms with physical (PHY) layer-based authentication to ensure reliable communication while minimizing computational and communication overheads. The initial authentication phase leverages a traditional public key infrastructure (PKI)-based approach, employing elliptic curve cryptography (ECC) and digital certificates to verify the legitimacy of communicating devices. Simultaneously, it extracts unique hardware-level features such as carrier frequency offset (CFO) and quadrature skewness from the devices. These features are then used to train a machine learning (ML) model during an offline phase managed by a regional centralized authority (RCA). For re-authentication, the system re-extracts these PHY-layer features from incoming orthogonal frequency division multiplexing (OFDM) symbols and verifies the device identity in real-time using the trained ML classifier. This cross-layer strategy enables continuous, lightweight identity verification without the need to exchange and validate cryptographic signatures for each message, thereby reducing system overhead. The proposed scheme further enhances privacy through the use of encrypted, frequently refreshed pseudo-identities, ensuring unlinkability and resistance to identity tracking. A formal security analysis using Burrows–Abadi–Needham (BAN) logic demonstrates the scheme's robustness against various threats, including impersonation, man-in-the-middle (MitM), replay, and Sybil attacks. Owing to its lightweight design and high level of security, the proposed scheme is particularly suitable for real-time, resource-constrained hospital environments. These features underscore the scheme’s potential to significantly strengthen the security and reliability of healthcare communications in modern medical infrastructures.
\end{abstract}

\IEEEoverridecommandlockouts
\begin{keywords}
Carrier frequency offset, Cross-layer authentication, Health care information exchange (HIE), Machine learning, Quadrature skewness.
\end{keywords}

\IEEEpeerreviewmaketitle

% ===================
% # I. Introduction #
% ===================

\section{Introduction}\label{S1}
The rapid digitization of healthcare systems has significantly transformed the way patient information is managed, stored, and exchanged, establishing health information exchange (HIE) as a vital component of modern medical practice \cite{b11,b12}. HIE facilitates the seamless and secure sharing of essential medical data among healthcare providers, institutions, and an increasing number of connected devices, thereby enhancing the quality of patient care, reducing medical errors, and improving overall operational efficiency \cite{b11}. As healthcare ecosystems evolve into highly interconnected networks, spanning hospitals, clinics, and remote care environments, the secure and efficient exchange of information has become essential for timely clinical decision-making and the continuity of care across diverse settings \cite{b13}. However, this growing dependence on digital infrastructure also expands the attack surface for cyber threats, making the protection of sensitive patient data a critical challenge for the future of healthcare innovation \cite{b14}.

Historically, healthcare data exchange was a manual and time-consuming process, limited to paper-based records and localized systems that offered little support for real-time collaboration or scalability \cite{b14}. During the mid-$20^{th}$ century, patient information was recorded by hand, stored in physical files, and shared through slow and manual methods such as postal mail or in-person transfer \cite{b14}. The introduction of electronic health records (EHRs) in the late $20^{th}$ century marked a significant advancement, replacing physical documentation with digital systems and enabling more efficient data storage and retrieval \cite{b15}. By the early 2000s, the emergence of networked healthcare systems accelerated the development of HIE, driven by the need to integrate diverse EHR platforms and improve interoperability \cite{b16}. A pivotal milestone was the enactment of the health information technology for economic and clinical health (HITECH) Act in 2009 in the United States, which incentivized EHR adoption and established standards for secure and standardized data sharing \cite{b16}. Since then, HIE has evolved into a complex and dynamic ecosystem that incorporates advanced technologies such as connected medical devices, wearable sensors, cloud computing, and artificial intelligence-driven analytics \cite{b17}. This progression highlights the increasing complexity and interconnectedness of healthcare delivery in the $21^{st}$ century.

\begin{figure}[t!]
\centerline{\includegraphics[width=8.8cm]{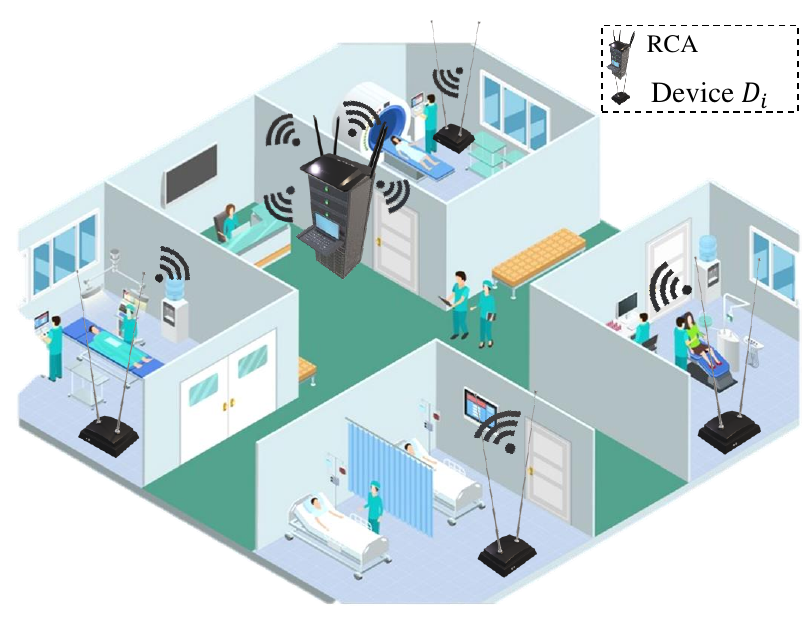}}
\setlength\abovecaptionskip{0.5\baselineskip}
\setlength\belowcaptionskip{-0.5cm}
\caption{System modeling.}
\label{f1}
\end{figure}

However, this technological advancement has introduced significant security challenges. The rapid proliferation of connected devices and the expansion of digital healthcare networks have substantially increased the attack surface, revealing vulnerabilities in legacy security mechanisms such as password-based authentication and basic encryption protocols \cite{b18}. Medical devices are particularly vulnerable due to their integration into networked environments, where they may serve as potential entry points for cyberattacks \cite{b22}. Similarly, the emergence of the internet-of-medical-things (IoMT) presents additional security and privacy concerns, as many of these devices lack robust built-in protection mechanisms \cite{b23}. Traditional security approaches, which were once adequate for isolated or less complex systems, are no longer sufficient for securing modern, resource-constrained hospital networks \cite{b19}. Advanced cyberattacks such as impersonation, man-in-the-middle (MitM), replay, and data tampering-exploit these limitations, threatening the confidentiality, integrity, and availability of patient data \cite{b20}. The consequences of such breaches can be severe, leading to compromised patient privacy, disruption of healthcare services, and potentially life-threatening delays in medical treatment \cite{b21}. As healthcare systems necessitate balance accessibility with security, there is a growing need for innovative and resilient solutions that can protect sensitive information and ensure data integrity while supporting the operational demands of real-time clinical environments.

In response to these challenges, this work proposes a cross-layer authentication scheme specifically designed to enhance the security of HIE networks within intra-hospital communication networks. The proposed approach combines advanced cryptographic methods with physical (PHY)-layer security features to provide a multi-layered defense against a wide range of cyber threats. The scheme begins with a centralized authority responsible for device registration and initial handshake authentication, utilizing elliptic curve cryptography (ECC) and digital certificates to establish a secure trust framework among communicating entities. To support ongoing secure communication, a lightweight re-authentication mechanism is introduced, which relies on subcarrier frequency offset (CFO) and skewness features extracted from orthogonal frequency division multiplexing (OFDM) symbols using the van-de-Beek algorithm \cite{b1}. These PHY-layer features serve as unique hardware-level fingerprints, enabling continuous identity verification and rapid detection of spoofed or unauthorized devices within the network.

Building on this foundation, the proposed scheme incorporates an additional layer of innovation by integrating context-aware adaptive security policies. Acknowledging the heterogeneity of devices and the varying sensitivity of data within hospital environments, the approach dynamically adjusts authentication requirements based on situational factors such as device type, data criticality, and network conditions. For example, high-priority transmissions involving real-time patient telemetry may trigger enhanced cryptographic verification, whereas routine administrative data exchanges employ streamlined processes to optimize resource utilization. To further enhance privacy, the scheme utilizes encrypted pseudo-identities that are periodically refreshed, preventing identity linkage across communication sessions and preserving patient anonymity. Security analysis demonstrates that this multi-faceted design provides robust protection against a broad range of attacks, including impersonation, man-in-the-middle (MitM), replay, and Sybil attacks, while maintaining low computational and communication overhead. This efficiency supports practical deployment in resource-constrained, real-time hospital networks. Generally, the main contributions of this paper are summarized as follows.

\begin{enumerate}
    \item This paper proposes a cross-layer authentication scheme for HIE networks that combines ECC-based PKI with PHY-layer features (e.g., carrier frequency offset and quadrature skewness) for lightweight and efficient device authentication.
    \item In addition, we introduce a machine learning-driven re-authentication mechanism using unique hardware-level PHY features extracted from OFDM signals, enabling real-time, resource-efficient identity verification without constant cryptographic overhead.
    \item Finally, the security of the proposed scheme is validated through formal analysis using Burrows–Abadi– Needham (BAN) logic, demonstrating strong resilience against common attacks, while achieving lower computational and communication overhead compared to traditional authentication methods, making it suitable for real-time healthcare environments.

\end{enumerate}

The remainder of this paper is organized as follows: Section~\ref{S2} reviews related works. Section~\ref{S3} describes the system model and relevant preliminaries. Section~\ref{S4} outlines the proposed authentication scheme in detail. Section~\ref{S5} presents the performance evaluation. Finally, Section~\ref{S7} concludes this work. For clarity, the acronyms used throughout this paper are summarized in Table \ref{tab:acronyms}.

\begin{table}[t]
\centering
\caption{List of acronyms}
\label{tab:acronyms}
\resizebox{0.485\textwidth}{!}{
\begin{tabular}{l|l}
\hline
\textbf{Acronym} & \textbf{Full Form} \\
\hline
BAN & Burrows-Abadi-Needham \\
CA & Centralized Authority \\
CFO & Carrier Frequency Offset \\
CP & Cyclic Prefix \\
ECC & Elliptic Curve Cryptography \\
EHR & Electronic Health Record \\
FFT & Fast Fourier Transform \\
GDPR & General Data Protection Regulation \\
HIE & Healthcare Information Exchange \\
HIPAA & Health Insurance Portability and Accountability Act\\
ML & Machine Learning \\
OID & Original Identity \\
OFDM & Orthogonal Frequency Division Multiplexing \\
PID & Pseudo-Identity \\
PKI & Public Key Infrastructure \\
RCA & Regional Centralized Authority \\
Rx & Receiver \\
SNR & Signal-to-Noise Ratio \\
SVM & Support Vector Machine \\
TPD & Tamper-Proof Device \\
Tx & Transmitter \\
\hline
\end{tabular}}
\end{table}

\section{Related Works}\label{S2}
Traditional security methods are no longer reliable in modern healthcare systems. Password-based authentication, in particular, is easily compromised by advanced computational techniques \citep{R2}. As a result, researchers develop more advanced authentication schemes that are specifically designed to address the constraints of healthcare environments \citep{R3}. This section reviews existing authentication approaches relevant to HIE networks, including multi-factor authentication (MFA), cryptographic and blockchain-based methods, physical-layer security techniques, and ML-driven models. MFA serves as a foundational element of healthcare security, addressing the inherent weaknesses of single-factor authentication. Sadhukhan et al. \cite{R4} propose a MFA framework for healthcare IoT systems that integrates ECC, biometrics, and smart devices to enhance the security of patient data access. Salim et al. \cite{R6} introduce a lightweight MFA scheme tailored for IoT-based healthcare environments, emphasizing computational efficiency for resource-constrained devices. A comprehensive review by Suleski et al. \cite{R7} evaluates existing MFA techniques within the internet-of-healthcare-things (IoHT), underscoring the need for adaptive frameworks capable of adjusting to varying risk levels while maintaining low-latency performance. Despite their advantages, traditional MFA solutions often require active user participation, which may introduce delays and inefficiencies in scenarios involving seamless device-to-device communication within intra-hospital networks \cite{R8}. The proposed scheme addresses these limitations by automating the authentication process and leveraging device-level physical features for real-time, continuous identity verification.

The IoMT enables connectivity through medical devices, necessitating robust authentication \cite{R9}. Cryptographic schemes based on PKI are prevalent \cite{R10}. Qi-Xie et al. \cite{b22} proposed an ECC-based authentication scheme within a blockchain-integrated IoMT framework, focusing on privacy-preserving communication. However, such schemes incur high computational overheads, limiting their use in resource-constrained devices. Blockchain-based authentication, as explored by Qi-Xie et al. \cite{b22}, uses private blockchains for data integrity but struggles with known key attacks. Jiang et al. \cite{R11} introduced a blockchain-based protocol for IoMT, improving scalability but requiring significant network resources. Lightweight cryptographic protocols, such as \cite{R12, R13}, reduce overheads but may compromise security robustness. Cross-layer authentication, integrating security across multiple protocol layers, offers a promising approach for securing HIE networks. PHY-layer security leverages unique wireless channel characteristics, CFO andI/Q modulator imbalance, as device-specific fingerprints for continuous authentication \cite{R14}. Zhang et al. highlighted its low overhead and suitability for IoMT systems \cite{R14}. Xu et al. proposed a PHY-layer authentication scheme using channel state information, achieving high accuracy with minimal cost \cite{R15}. Lee et al. explored cross-layer authentication in IoT networks, combining physical-layer features with network-layer cryptography, but its healthcare application is limited \cite{R16}. Recent advancements by Tian et al. \cite{R17} introduced a cross-layer scheme using radio frequency (RF) fingerprints, improving resilience against spoofing but requiring complex signal processing. Zhang et al. \cite{R18} developed a PHY-layer authentication protocol for healthcare IoT, emphasizing energy efficiency.

ML significantly enhances authentication accuracy and adaptability in healthcare systems by enabling intelligent classification of PHY-layer features to differentiate between legitimate devices, thus supporting lightweight reauthentication \cite{R19}. Almaiah et al. \cite{b24} demonstrate the application of ML within a hybrid authentication model for the IoHT, achieving high accuracy with minimal computational overhead. Kumar et al. \cite{b25} propose an improved Elman neural network (IENN) to assess data sensitivity levels within an IoMT framework, emphasizing security for smart healthcare applications. Deep learning approaches, such as the RF fingerprint authentication method introduced by Huang et al. \cite{R20}, deliver high classification precision but demand substantial computational resources. Alsalman et al. \cite{R21} develop an ML-based intrusion detection system for IoMT networks, effectively enhancing anomaly detection and network resilience. Additionally, Nguyen et al. \cite{R22} explore the use of federated learning to enable collaborative model training across distributed healthcare devices while preserving patient privacy; however, this approach incurs increased communication overheads. The proposed scheme addresses these trade-offs by integrating a lightweight neural network for PHY-layer feature classification, offering a practical balance between accuracy, efficiency, and scalability in resource-constrained hospital environments.

\section{System Modelling and Preliminaries}\label{S3}
This section presents the system model and the fundamental preliminaries employed in this study.
\subsection{System modelling}\label{S3.1}
The network comprises the following entities, see Fig. \ref{f1}.
 \subsubsection{Centralised authority ($CA$)}\label{S4.1.1}
The $CA$ is a third party that plays a crucial role in the system by serving as a trusted central entity responsible for overseeing the generation and management of the public system parameters. These parameters include cryptographic keys and encryption functions, which are essential for secure communication within the network. The $CA$ is also responsible for initialising the system, registering the regional centralised authorities ($RCA$s) and several communicating nodes/devices (\(D_i\)), and ensuring the overall security and integrity of the communication process.

 \subsubsection{Regional centralised authority ($RCA$)}\label{S4.1} 
$RCA$s are infrastructure components deployed on each floor or department that manage and access a number of communicating nodes in the network. They serve as communication hubs, interacting with the communicating nodes. The $RCA$ plays a key role in facilitating secure communication by verifying the authenticity of messages and ensuring that communication is conducted securely and efficiently. 

 \subsubsection{Communicating devices \(D_i\)}\label{S4.1} 
Each $D_i$ in the network is deployed in every section or room of the floor or department, enabling medical devices to process and transmit patient scanning images, including CT scans, X-rays, ECGs, among others. These messages must be sent to the administration office through the $RCA$ in accordance with specific security and privacy requirements, defined as follows. 
\begin{enumerate}
    \item \textit{Confidentiality}: Ensuring that sensitive information is accessible only to authorised parties.
    \item \textit{Privacy preservation}: Safeguarding the personal information of patients.
    \item \textit{Unlinkability}: Ensuring that transmitted messages cannot be linked to specific individuals.
    \item \textit{Traceability}: Allowing the identification of message senders or recipients if necessary.
    \item \textit{Security against impersonation, modification, and replay attacks}: protecting the system from unauthorised alterations or attempts to replay or mimic messages. 
\end{enumerate}

\subsection{Preliminaries}
This part introduces the key functions incorporated into the proposed authentication scheme.
\subsubsection{Carrier frequency offset estimation (\textit{CFO\_Extractor}) per Hz}\label{S3.2} CFO arises due to inherent hardware imperfections in the local oscillators of different wireless devices. Each transmitter introduces a distinct frequency offset, leading to a phase shift between the carrier frequencies of the transmitter ($Tx$) and receiver ($Rx$). These variations in CFO can serve as unique fingerprints for device identification or classification. While CFO may exhibit slight variations over time, its overall pattern remains consistent for a given device. In this study, CFO estimation is performed at the $Rx$ side using the method proposed by Van de Beek \cite{b1}. The Van de Beek algorithm is a widely adopted approach for CFO estimation in OFDM systems. It leverages the cyclic prefix (CP) within OFDM symbols to determine frequency offsets across different devices. The core principle of this method involves autocorrelation between repeated symbols to detect frequency shifts. The CFO estimation equation is given by:
\begin{equation}\label{e1}
\hat{\epsilon} = \frac{1}{2\pi L} \arg \left( \sum_{n=0}^{N-L-1} r[n] r^*[n+L] \right)
\end{equation}
where \( \hat{\epsilon} \) represents the normalised CFO estimate, while \( r[n] \) denotes the received signal samples. The symbol repetition interval is given by \( L \), which is often equivalent to the cyclic prefix length, and \( N \) represents the total number of samples. Additionally, the function \( \arg(\cdot) \) extracts the phase angle from the complex correlation result. To obtain the actual CFO in hertz, the normalised CFO estimate \( \hat{\epsilon} \) must be converted using the sampling frequency \( F_s \) and a scaling factor \( N \), which may represent the number of symbols. The estimated frequency offset ($\widehat{C F O}$) in Hz can be expressed as:  
\begin{equation}\label{e2}
\widehat{C F O} = N \cdot \hat{\epsilon} \cdot F_s
\end{equation}
For more details, refer to \cite{b1}.

\subsubsection{Quadrature skewness estimation (\textit{Quad\_Skew}) per degrees}\label{s3.3}
The quadrature skewness angle \( \theta_{\text{skew}} \) measures how much the I/Q coordinate system deviates from the ideal orthogonal $X/Y$ coordinate system. Mathematically, this can be expressed as:
\begin{equation}\label{e3}
    \widehat{\theta}_{\text{skew}} = \tan^{-1} \left( \frac{\sum I[n] Q[n]}{\sum I[n]^2 - \sum Q[n]^2} \right)
\end{equation}
where \( I[n] \) and \( Q[n] \) denote the In-phase and Quadrature components, respectively. The numerator, \( \sum I[n] Q[n] \), quantifies the correlation between these components, which ideally should be zero in a perfectly balanced system. The denominator, \( \sum I[n]^2 - \sum Q[n]^2 \), accounts for any power asymmetry between the two signals, providing a normalisation factor that reflects deviations from the ideal orthogonality and equal power distribution in the I/Q plane. The result is an angle in degrees (after converting from radians), indicating how much the I/Q axes have rotated from their ideal positions. The interpretation of \( \theta_{\text{skew}} \):
When \( \theta_{\text{skew}} = 0^\circ \), the I/Q axes remain perfectly aligned with the $X$ and $Y$ coordinate system, indicating no quadrature skew. A positive skew angle (\( \theta_{\text{skew}} > 0^\circ \)) corresponds to a counterclockwise rotation of the $Q$ axis relative to the $Y$-axis, while a negative skew angle (\( \theta_{\text{skew}} < 0^\circ \)) signifies a clockwise rotation. Larger skew angles indicate a more pronounced misalignment, which can adversely impact demodulation performance in OFDM receivers by introducing phase distortions and impairing symbol detection accuracy. 
For more details, refer to \cite{b2}.

This study proposes using two functions, \textit{CFO\_Extractor} and \textit{Quad\_Skew}, to extract the unique impairments inherent in the $Tx$s' characteristics. These extracted features can serve as distinctive identifiers, enabling differentiation among all communicating devices within the network.
\begin{figure*}
 \centerline{\includegraphics[width=17cm]{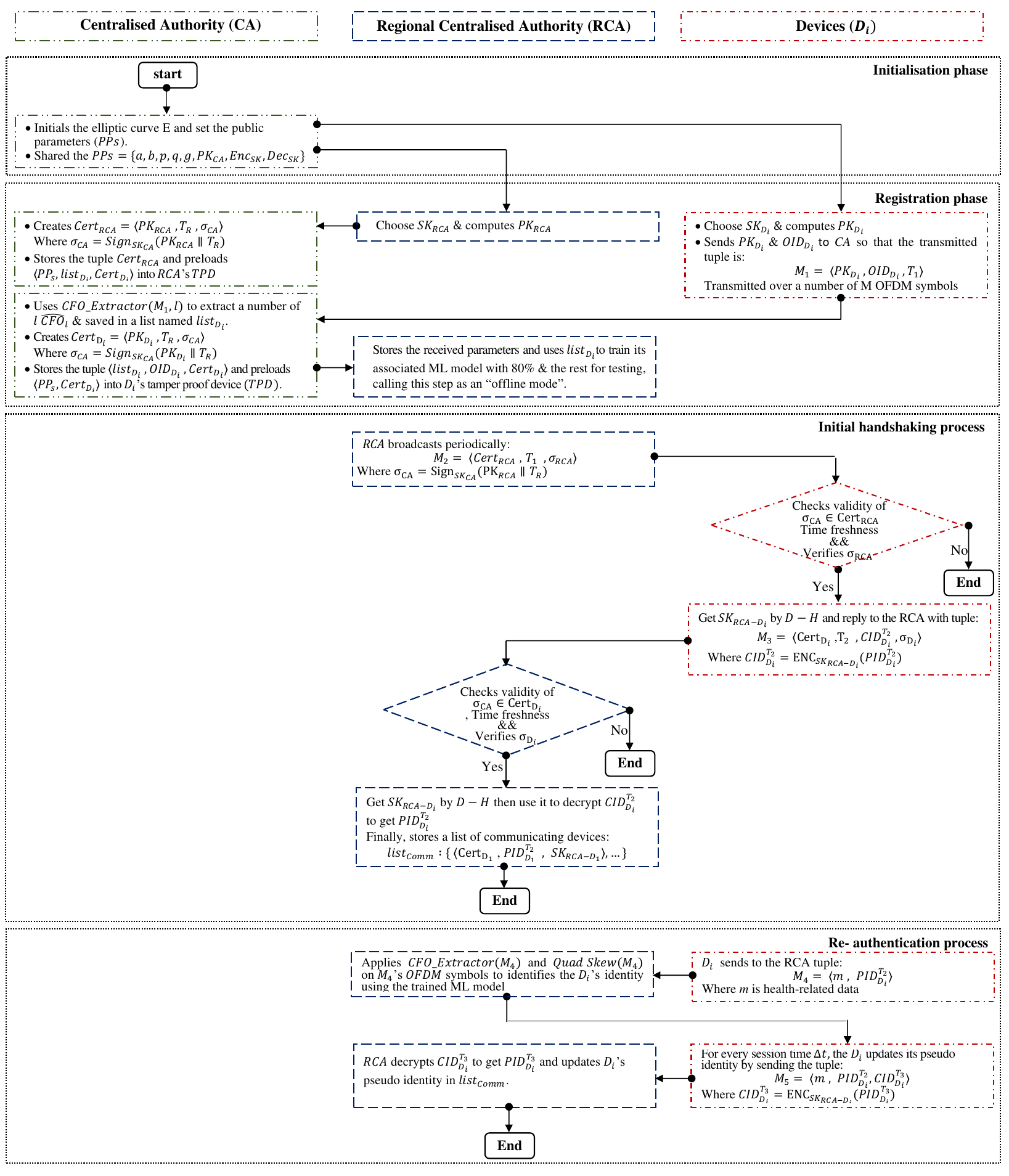}}
\setlength\abovecaptionskip{0.5\baselineskip}
\setlength\belowcaptionskip{-0.5cm}
\caption{Initial handover authentication and re-authentication process.}
\label{f2} 
\end{figure*}
\begin{table}[]
    
     \caption{The recommended domain parameters of the 160-bit elliptic curve ``secp160k1'' in the hexadecimal \cite{b222}}
    \label{tab:my_recom}
    \resizebox{.48\textwidth}{!}{
    \begin{tabular}{c|c}
      \hline
       \textbf{Par.}  & \textbf{Recommended value} \\
       \hline
       $a$  & $00000000$ $00000000$ $00000000$ $00000000$ $00000000$\\
       $b$& $00000000$ $00000000$ $00000000$ $00000000$ $00000007$\\
      $p$& $FFFFFFFF$ $FFFFFFFF$ $FFFFFFFF$ $FFFFFFFF$ $FFFFAC73$\\
       \hline
       \multirow{2}{.2cm}{$g$}& $04$ $3B4C382C$ $E37AA192$ $A4019E76$ $3036F4F5$ $DD4D7EBB$\\
       &$938CF935$ $318FDCED$ $6BC28286$ $531733C3$ $F03C4FEE$\\
       \hline
       $q$& $01$ $00000000$ $00000000$ $0001B8FA$ $16DFAB9A$ $CA16B6B3$\\
       \hline
    \end{tabular}
    }
   
\end{table}

\section{Scheme Modelling}\label{S4}
The proposed scheme is structured into three phases: Initialisation, registration and authentication, and re-authentication, as detailed below.
\subsection{Initialisation phase}\label{S4.2}The $CA$ initiates the generation of the system's public parameters, which are subsequently shared by participants to streamline the registration processes of other $RCAs$ and communicating nodes $(D_i)$. The details of the $CA$ initialisation phase are as follows:
\begin{itemize}
    \item The $CA$ chooses two prime numbers, $p$ and $q$ where $p$ is a large prime number, used to initialise the elliptic curve $E: y^2=x^3+ax+b\ mod\ p$ where $(a,b)\in Z_q^\ast$. We use the recommended domain parameters in \cite{b222} of the 160-bit elliptic curve ‘‘secp160k1’’, for 80-bit security, see Table \ref{tab:my_recom}.
\end{itemize}
\begin{itemize}
    \item The $CA$ chooses the generator $g$ with a length of $q$ and establishes the cyclic additive group $G$ , combines all points on $E$.
\end{itemize}
\begin{itemize}
    \item The $CA$ selects at random a private key ${SK}_{CA}\in{Z}_{q}^\ast$ then computes its public key ${PK}_{CA}= {SK}_{CA}\cdot g$.
\end{itemize}	
\begin{itemize}
    \item Finally, \( CA \) chooses symmetric encryption and decryption functions, denoted as \( {Enc}_{SK}(.) \) and \( {Dec}_{SK}(.) \), respectively so that the overall system's publicly accessible parameters are given by:
    \[ PPs = \{a, b, p, q, g, {PK}_{CA}, {Enc}_{SK}, {Dec}_{SK} \} \].
\end{itemize}	
\subsection{Registration phase}\label{S4.2} In this phase, the CA registers each RCA in a specific region and its associated devices within its communication range. Accordingly, this phase comprises two stages, which are defined as follows:
\begin{itemize}

    \item \textit{Stage 1 ($D_i$ registration)}: This phase comprises the following steps:
        \begin{enumerate}
            \item \( D_i \) chooses a random integer \( SK_{D_i} \in \mathbb{Z}_q^\ast \) as its private key, then computes its corresponding public key as \( PK_{D_i} = SK_{D_i} \cdot g \).
            \item \( D_i \) sends \( PK_{D_i} \) and its original identity \( OID_{D_i} \) to the \( CA \), so that the transmitted tuple is \( M_1 = \langle PK_{D_i}, OID_{D_i}, T_1 \rangle \). This tuple can be transmitted over a number of \( M \) OFDM symbols.
            \item Accordingly, the \( CA \) uses the function \linebreak \( CFO\_Extractor(M_{1, l}) \) to extract a number of \( l \) \( \widehat{CFO}_l \) estimates, \( \forall l \in \{1, \dots, M\} \), which are saved in a list named \( list_{D_i} \).            
            \item The $CA$ creates $D_i$'s long-term digital certificate \ $Cert_{D_i}=\langle{PK}_{D_i}, T_R,\sigma_{CA}\rangle$, where $T_R$ and $\sigma_{CA}=Sign_{SK_{CA}}({PK}_{D_i} \| T_R)$ represent $Cert_{D_i}$'s expiry date and $CA$'s digital signature, respectively.
            \item At last, the $CA$ stores the tuple $\langle list_{D_i}, $ $ {OID}_{D_i}, $ $ Cert_{D_i}\rangle$ and preloads $\langle PPs, Cert_{D_i} \rangle$ into $D_i$'s tamper proof device (TPD).
        \end{enumerate}

    \item \textit{Stage 2 ($RCA$ registration)}: In this stage the $CA$ performs the following steps:
        \begin{enumerate}
     \item The $RCA$ chooses a random integer ${SK}_{RCA}\in Z_q^\ast$ as its private key then computes its corresponding public-key ${PK}_{RCA}={{SK}}_{RCA} \cdot {g}$.
     \item The $CA$ creates $RCA$'s long-term digital certificate \ $Cert_{RCA}=\langle{PK}_{RCA}, T_R,\sigma_{CA}\rangle$, where $T_R$ and $\sigma_{CA}=Sign_{SK_{CA}}({PK}_{RCA} \| T_R)$ represent $Cert_{RCA}$'s expiry date and $CA$'s digital signature, respectively.
    \item The $CA$ stores the $Cert_{RCA}$ and preloads the tuple $\langle PPs, list_{D_i}, Cert_{RCA} \rangle$, $\forall i=\{1, \dots, N\}$, into $RCA$'s TPD, where $N$ is the total number of devices located in the $RCA$'s communication range.
    \item Finally, the $RCA$ stores the received parameters and uses $list_{D_i}$, $\forall i=\{1, \dots, N\}$, to train its associated machine learning model with $80\%$ from the overall dataset and the rest of the dataset is used for testing, calling this step as an ``offline mode''. 
    \end{enumerate}
\end{itemize}

\subsection{Initial handshaking and re-authentication phase}: This phase is performed on each floor or within each department between the RCA and the communicating devices. This phase consists of two processes, which are defined as follows.
\subsubsection{Initial handshaking process}
Generally, this process comprises the following steps.
        \begin{enumerate}
            \item The \( RCA \) periodically broadcasts $M_2 =$ $ \langle Cert_{RCA},$ $ T_1,$ $ \sigma_{RCA} \rangle $, where \( \sigma_{RCA} \) is the \( RCA \)'s digital signature, generated at the timestamp \( T_1 \).
            \item The $D_i$ in turn checks the validity of \( \sigma_{CA} \in Cert_{RCA} \), ensures the time freshness of \( T_1 \), and verifies \( \sigma_{RCA} \). Then, it derives the shared key through the Diffie-Hellman exchange as \(SK_{RCA-D_i} = SK_{D_i} . PK_{RCA} \). Finally, the $D_i$ replies to the $RCA$ by sending the tuple
            $M_3 = \langle Cert_{D_i}, T_2, CID_{D_i}^{T_2}, \sigma_{D_i} \rangle$, where \( CID_{D_i}^{T_2} = ENC_{SK_{RCA-D_i}}(PID_{D_i}^{T_2}) \) represents the encryption of the pseudo-identity \( PID_{D_i}^{T_2} \) at timestamp \( T_2 \) using the shared key \( SK_{RCA-D_i} \).
            \item The $RCA$ checks the validity of \( \sigma_{CA} \in Cert_{D_i} \), ensures the time freshness of \( T_2 \), and verifies \( \sigma_{D_i} \). Then, it derives the shared key through the Diffie-Hellman exchange as \(SK_{RCA-D_i} = SK_{RCA} . PK_{D_i} \), used to decrypt $CID_{D_i}^{T_2}$ to get $PID_{D_i}^{T_2}$. Finally, the $RCA$ stores a list of the communicating devices $list_{Comm}: \{\langle {Cert}_{D_i} ,{PID}_{D_i}^{T_2} , {SK}_{RCA-D_i} \rangle, \cdots \}$, $\forall i=\{1, \cdots, N\}$.  
        \end{enumerate}

\subsubsection{The re-authentication process}
This process is performed for every broadcasted health-related data ($m$) from $D_i$ to the $RCA$ in its communication range. Generally, this process involves the following steps:
        \begin{enumerate}
            \item The $D_i$ sends the tuple $M_4 = \langle m, {PID}_{D_i}^{T_2} \rangle$ to the $RCA$. Accordingly, the $RCA$ applies \textit{CFO\_Extractor} $(M_4)$ and \textit{Quad\_Skew}($M_4$) functions on $M_4$'s OFDM symbols to extract $\widehat{C F O}_{D_i}$ and $\widehat{\theta}_{\text{skew}(D_i)}$ using equations \eqref{e2} and \eqref{e3}, respectively. After that, the $RCA$ identifies the $D_i$'s identity using the trained ML model, a process referred to as the online mode.

            \item For every session time $\Delta t$, the $D_i$ updates its pseudo identity by sending the tuple $M_5 = \langle m, {PID}_{D_i}^{T_2}, {CID}_{D_i}^{T_3} \rangle$ to the $RCA$, where \( CID_{D_i}^{T_3} = ENC_{SK_{RCA-D_i}}(PID_{D_i}^{T_3})\). 
            \item The $RCA$ in turn decrypts $CID_{D_i}^{T_3}$ to get $PID_{D_i}^{T_3}$ and updates $D_i$'s pseudo identity in $list_{Comm}$. This way ensures the unlinkability between the broadcasted messages from $D_i$ in different sessions.
        \end{enumerate}

\section{Performance Evaluation}\label{S5}
This section presents informal and formal security analyses of the proposed method, discusses the performance of the PHY-layer re-authentication process, and evaluates its computational and communication costs.

\subsection{Security analysis}
This part analyzes the proposed scheme's security and privacy through both formal and informal methods.

\subsubsection{BAN-logic formal security analysis}\label{S6.1}
The Burrows–Abadi–Needham logic provides a formal framework for analyzing authentication protocols \cite{b3} by modeling the beliefs and knowledge of the participating entities. By representing protocol steps and assumptions within a logical system, BAN logic enables the rigorous verification of security properties, such as mutual authentication, freshness, and resistance to against common threats, including replay, man-in-the-middle, and impersonation attacks. Table \ref{tab:ban-notations} summarizes the notations used in the BAN logic proof.

\begin{table}[t!]
\centering
\caption{{BAN logic notations}}
\resizebox{.48\textwidth}{!}{\label{tab:ban-notations}
\vspace{0.5em}
\begin{tabular}{l|l}
\hline
\textbf{Symbol} & \textbf{Description} \\
\hline
$P, Q$ & Communicating entities, e.g. CA, RCA, $D_i$ \\
$K$ & Cryptographic shared key \\
$\{X\}_K$ & Message $X$ encrypted with key $K$ \\
$\{X\}_{K_{PHY}}$ & Message $X$ encrypted with PHY-features key $\{X\}_{K_{PHY}}$ \\
$P \mid\equiv X$ & Principal $P$ believes $X$ \\
$P \triangleleft X$ & Principal $P$ sees message $X$ \\
$P \mid\sim X$ & Principal $P$ once said $X$ \\
$\#(X)$ & Message $X$ is fresh \\
$P \mid\Rightarrow X$ & Principal $P$ has jurisdiction over $X$ \\
$P \stackrel{K}{\leftrightarrow} Q$ & $P$ and $Q$ share the symmetric key $K$ \\
%$P \Rightarrow X$ & Principle has control over $X$\\
$P \xrightarrow{K} Q$ & $K$ is the public key of $P$. \\
\hline
\end{tabular}}
\end{table}

\paragraph{Rules}: A set of deductive rules is applied to the protocol messages and initial beliefs. These rules enable the principals to infer new beliefs from received messages. Table \ref{tab:ban-rules} lists and defines these rules.
\begin{table*}
\centering
\caption{ {The  rules  listed  in  the  BAN-logic  analysis}}
\label{tab:ban-rules}
\vspace{0.5em}
\resizebox{1\textwidth}{!}{\begin{tabular}{ c|c|c|c }
\hline
\textbf{No.} & \textbf{Rule} & \textbf{Description} &\textbf{Definition}\\
\hline
{R1} &  {Message Meaning Rule} & 
 $ \frac{P \mid\equiv (P \stackrel{K}{\leftrightarrow} Q ) , P \triangleleft \{X\}_K} {P \mid\equiv Q \mid\sim X}$  & If $P$ believes in the shared key $K$ and sees a message $X$ encrypted with $K$,\\
 &for a shared key&&then $P$ can conclude that $Q$ once said $X$\\

{R2} & {Message Meaning Rule} & 
 $ \frac{P \mid\equiv (Q \stackrel{K}{\rightarrow} P ) , P \triangleleft \{X\}_{K^{-1}}} {P \mid\equiv Q \mid\sim X}$  & If $P$ believes in the public key $K$ and sees a message $X$ encrypted with $Q's$ \\
 &for a public key&&private key, then $P$ can conclude that $Q$ once said $X$\\

{R3} & {Nonce Verification Rule} & 
$\frac{P \mid\equiv \#(X) , (P \mid\equiv Q \mid\sim X)}{P \mid\equiv (Q \mid\equiv X)}$ & If $P$ believes $X$ is fresh, and $P$ believes $Q$ once said $X$, then $P$ believes $Q$\\
&&& believes $X$  \\

{R4} & {Jurisdiction Rule} & 
$\frac{P \mid\equiv (Q \mid\Rightarrow X) , P \mid\equiv (Q \mid\equiv X)}{P \mid\equiv X}$ & If $P$ believes $Q$ has jurisdiction over $X$, and $P$ believes $Q$ believes $X$, then \\
&&&$P$ believes $X$  \\

{R5} & {Freshness Rule} & 
$\frac{P \mid\equiv \#(X)}{P\mid\equiv \#(X,Y)  }$ & f $P$ believes $X$ is fresh, then $P$ believes any part of formula is fresh\\

\hline
\end{tabular}}
\end{table*}

\paragraph{Idealized protocol messages}: The protocol messages are abstracted into the following idealized forms:
\begin{itemize}
    \item $S_1$ : $RCA$ $\rightarrow$ \text{$D_i$} :$\{{Cert_{RCA}}, T_1\}_{{SK}_{RCA}}$  \\where \[Cert_{RCA}=\{{PK}_{RCA}, T_R\}_{SK_{CA}}\]
 
\item $S_2$ : ${D_i}$ $\rightarrow$ \text{$RCA$}: $\{  Cert_{D_i}, T_2, CID_{D_i}^{T_2}\}_{SK_{D_i}} $ \\where:
\[Cert_{D_i}=\{{PK}_{D_i}, T_R\}_{SK_{CA}}\]   \[CID_{D_i}^{T_2} =\{PID_{D_i}^{T_2}\}_{SK_{RCA-D_i}}\]

\item $S_3$ : $D_i$ $\rightarrow$ \text{$RCA$} : \{$m$, 
 $PID_{D_i}^{T_2}$\}$_{K_{PHY}}$ 

\item $S_4$ : $D_i$ $\rightarrow$ \text{$RCA$} : \{$m$, $PID_{D_i}^{T_2}$, $CID_{D_i}^{T_3}$ \}$_{K_{PHY}}$ 
\\where:
\[CID_{D_i}^{T_3} = \{PID_{D_i}^{T_3}\}_{SK_{RCA-D_i}}\]

\end{itemize}

\paragraph{Goals}: The objective of the BAN logic analysis is to verify that the proposed scheme satisfies the following security properties:

\begin{itemize}
    \item \textit{Goal 1}: \ {$D_i \mid\equiv (D_i \stackrel{SK_{RCA-D_i}}{\leftrightarrow} {RCA})$} \\
    \textit{Meaning:} $D_i$ is convinced that it shares a secure key with the $RCA$.

    \item \textit{Goal 2}: \ {$D_i \mid\equiv (RCA \mid\equiv \ S_1)$} \\
    \textit{Meaning:} $D_i$ trusts that the initial handshaking broadcasted message $S_1$ originated from the $RCA$.
   
    \item \textit{Goal 3}: \ {$RCA\mid\equiv ({RCA} \stackrel{SK_{RCA-D_i}}{\leftrightarrow} D_i$)} \\
    \textit{Meaning:} The $RCA$ is assured that a shared key $SK_{RCA-D_i}$ established between itself and $D_i$.

\item \textit{Goal 4}: \textbf{$RCA\mid\equiv (D_i \ \mid\equiv \ S_2)$} \\
    \textit{Meaning:} The $RCA$ trusts that the  message $S_2$ originated from $D_i$.
    
    \item \textit{Goal 5}: \textbf{$RCA\mid\equiv (D_i \stackrel{K_{PHY}} {\rightarrow} \ RCA)$} \\
    \textit{Meaning:} The $RCA$ trusts  the  PHY-features key of $D_i$.

    \item \textit{Goal 6}: \ {$RCA \mid\equiv S_3$} \\
    \textit{Meaning:} The $RCA$ ultimately accepts the authentication message $S_3$ (which includes the encrypted pseudo-identity) as authentic, thereby binding $D_i$ to its transmitted identity.
    
    \item \textit{Goal 7}: \ {$RCA \mid\equiv S_4$} \\
    \textit{Meaning:} The $RCA$ ultimately accepts the authentication message $S_4$ (which includes the encrypted new pseudo-identity) as authentic, thereby ensuring the freshness and authenticity of the updated pseudo-identity.
\end{itemize}

\paragraph{Assumptions}: The security proof relies on the following basic assumptions (initial beliefs) that are presumed true prior to protocol execution:

\begin{enumerate}
    \item $A_1$ [{$D_i$ $\mid\equiv$} $\#(T_1)$]: The $D_i$ assumes that the timestamp $T_1$ is fresh.

    \item $A_2$ [$RCA$ $\mid\equiv$ $\#(T_2)$]: The device $RCA$ considers $T_2$ to be fresh.

    \item $A_3$ [$RCA$ $\mid\equiv$ $\#(T_3)$]: The $RCA$ regards $T_3$ as ensuring the freshness of $S_4$.

    \item $A_4$ [{$RCA$ $\mid\equiv$} (\text{$CA$}  
 $\stackrel{PK_{CA}}{\rightarrow}$ $RCA$ )]: The $RCA$ trusts that the $CA$ is authoritative in certifying public keys.

   \item $A_5$ [{$D_i$ $\mid\equiv$} (\text{$CA$}  
 $\stackrel{PK_{CA}}{\rightarrow}$ $D_i$ )]: $D_i$ trusts the $CA’s$ certification of public keys.

\item  $A_6$ $\left[ \frac{\textit{RCA} \ \mid\equiv \textit\ (CA \stackrel{PK_{CA}}{\rightarrow} RCA ) ,\ RCA \triangleleft \{ PK_{D_i},\ T_R\}_{SK_{CA}}} {RCA \ \mid\equiv \ (\text{$D_i$} \stackrel{PK_{D_i}}{\rightarrow} RCA)}\right]$: If $RCA$ trusts that the $CA$ is authoritative in certifying public keys and receive a message signed with $CA$ private key, then $RCA$ trusts the device’s $D_i$ public key as issued by the trusted $CA$.

    \item $A_7$: $\left[\frac{\textit D_i \ \mid\equiv \textit \ (CA \stackrel{PK_{CA}}{\rightarrow} D_i ) , \ \textit D_i \triangleleft \{ \textit PK_{RCA}, \ \textit T_R\}_{SK_{CA}}} {D_i \ \mid\equiv \ (\text{$RCA$} \stackrel{PK_{RCA}}{\rightarrow} D_i)}\right]$: If $D_i$ trusts that the $CA$ is authoritative in certifying public keys and receive a message signed with $CA$ private key, then $D_i$ trusts the device’s $RCA$’s public key as issued by the trusted $CA$.

\item  $A_8$ $\left[ \frac{\textit{RCA} \ \mid\equiv \textit\ (\text{$D_i$} \stackrel{PK_{D_i}}{\rightarrow} RCA) ,\ RCA \triangleleft \{    \{S_2\}_{K_{PHY}}\}_{SK_{D_i}} }   {RCA \ \mid\equiv \ (\text{$D_i$} \stackrel{K_{PHY}}{\rightarrow} RCA)}\right]$: If $RCA$ trusts the device’s $D_i$ public key as issued by the trusted $CA$ and receive a message with $K_{PHY}$ signed with $D_i$ private key, then $RCA$ trusts the device’s $D_i$ ${K_{PHY}}$ is it's PHY-features key as fingerprint.

    \item $A_9$: $\left[RCA \mid\equiv \ ({D_i} \Rightarrow S_3)\right]$: The $RCA$ trusts that the device $D_i$ is issued the message $S_3$.

      \item $A_{10}$: $\left[RCA \mid\equiv \ ({D_i} \Rightarrow S_4)\right]$: The $RCA$ trusts that the device $D_i$ is issued the message $S_4$.
\end{enumerate}

\paragraph{Implementation}: This subsection provides a detailed analysis of the proposed authentication scheme using the BAN logic framework. The analysis systematically evaluates the protocol messages exchanged between the $RCA$ and $D_i$, demonstrating how the scheme achieves its security goals through logical inference rules and initial assumptions. The protocol is proven to ensure mutual authentication, shared key establishment, and secure re-authentication, while remaining resilient against replay, man-in-the-middle, and impersonation attacks.

\begin{itemize}
    \item \textit{Step 1}: $RCA$ broadcasts initial handshaking message \( S_1 \), \( D_i \) receives the message \( S_1 \),
        applies $A_5$, $S_1$ to $A_7$, resulting in the following outcome: $O_1$ : ${D_i \ \mid\equiv \ (\text{$RCA$} \stackrel{PK_{RCA}}{\rightarrow} D_i)}$. Therefore \( D_i \) computes the shared key: \( SK_{RCA-D_i} = SK_{D_i} \cdot PK_{RCA} \) (using D-H protocol), leading to $O_2$ : {$D_i \mid\equiv (D_i \stackrel{SK_{RCA-D_i}}{\leftrightarrow} {RCA})$}, achieving \textit{Goal 1}. 
        \item \textit{Step 2}: By applying $O_1$ and \( S_1 \) to $R_2$ from Table \ref{tab:ban-rules}, the outcome is $O_3$ : \({D_i \mid\equiv RCA \mid\sim S_1}\). Next, by applying $A_1$ and $S_1$ to $R_5$ from Table \ref{tab:ban-rules}, we have $O_4$ :\( D_i \mid\equiv \#(S_1) \). Accordingly, by applying $O_3$ and $O_4$ to $R_3$ from Table \ref{tab:ban-rules}, we have \({D_i \mid\equiv (RCA \mid\equiv \ S_1)}\), satisfying \textit{Goal 2}.   

    \item \textit{Step 3}: Device responds with the message \( S_2 \), $RCA$ receives \( S_2 \). From assumption $A_4$ and $S_2$ applied in $A_6$,  infers $O_5$ : \( RCA \mid\equiv (D_i \stackrel{PK_{D_i}}{\rightarrow} RCA) \). Therefore \( RCA \) computes the shared key: \( SK_{RCA-D_i} = SK_{RCA} \cdot PK_{D_i} \) and have $O_6$ : {$RCA\mid\equiv ({RCA} \stackrel{SK_{RCA-D_i}}{\leftrightarrow} D_i)$}, achieving \textit{Goal 3}. 
        \item \textit{Step 4}: By applying $O_5$ and $S_2$ to $R_2$ from Table \ref{tab:ban-rules}, the outcome is $O_7$ : \({RCA \mid\equiv (D_i \mid\sim S_2)}\). Next, by applying $A_2$ and $S_2$ to $R_5$ from Table \ref{tab:ban-rules}, we have $O_8$ : \( RCA \mid\equiv \#(S_2) \). Accordingly, by applying $O_7$ and $O_8$ to $R_3$ from Table \ref{tab:ban-rules}. Thus, \( RCA \mid\equiv (D_i \mid\equiv S_2) \), achieving \textit{Goal 4}. Therefore, from $A_8$ we have outcome $O_9$ : \textbf{$RCA\mid\equiv (D_i \stackrel{K_{PHY}} {\rightarrow} \ RCA)$}. Achieving \textit{Goal 5}.
   
       \item \textit{Step 5}:  Re-authentication with message \( S_3 \). $RCA$ receives the message \( S_3 \), then identifies \( D_i \) using \( PID_{D_i}^{T_2} \), which is tied to \( SK_{RCA-D_i} \) established in $Step\ 3$. By applying $O_9$ and $S_3$ to $R_2$ from Table \ref{tab:ban-rules}, we have outcome $O_{10}$ : \( RCA \mid\equiv (D_i \mid\sim S_3) \). Since \( S_3 \) includes \( PID_{D_i}^{T_2} \) which it's in the same session at \( T_2 \)  checked to be fresh at $Step\ 4$, $O_{11}$ : \( RCA \mid\equiv \#(S_3) \). Applying $O_{10}$ and $O_{11}$ to $R_3$ from Table \ref{tab:ban-rules},  \( RCA \mid\equiv (D_i \mid\equiv S_3) \). With  $A_9$ applying to $R_4$ from Table \ref{tab:ban-rules}, $O_{12}$: \( RCA \mid\equiv S_3 \), achieving \textit{Goal 6}.
   \item \textit{Step 6}: New session is initialized with the message \( S_4 \). $RCA$ receives the message \( S_4 \).
        $RCA$ applies $O_6$ and $S_4$ to $R_1$ from Table \ref{tab:ban-rules}, thus $O_{13}$ : \( RCA \mid\equiv (D_i \mid\sim S_4) \).
        With assumption $A_3$ applying $R_5$ from Table \ref{tab:ban-rules}, $O_{14}$ : \( RCA \mid\equiv \#(S_4) \).
    Applying $O_{13}$ and $O_{14}$ to $R_3$ from Table \ref{tab:ban-rules}. Therefore, $O_{15}$ : \( RCA \mid\equiv (D_i \mid\equiv S_4) \). From $A_{10}$ and $O_{15}$ applying to $R_4$  from Table \ref{tab:ban-rules}, $O_{16}:$ \( RCA \mid\equiv S_4 \), achieving \textit{Goal 7}.
    \end{itemize}

The protocol ensures mutual authentication between $RCA$ and \( D_i \), establishes a secure shared key, and supports secure re-authentication and session refresh. By leveraging the BAN logic rules and assumptions, the scheme is formally verified to be robust against replay, man-in-the-middle, and impersonation attacks, making it suitable for secure communication in IoT environments.

\subsection{Informal Security Analysis}
This part demonstrates that the proposed scheme satisfies the security and privacy requirements outlined in Subsection \ref{S4.1}.

\subsubsection{Privacy preservation}
Ensuring the privacy of the communicating nodes $D_i$ is a critical aspect of secure communication systems. When privacy is upheld, an adversary cannot infer a node's identity from its responses, as only the $CA$ knows $OID_{D_i}$. The proposed scheme achieves strong privacy preservation through a public key certification process during the initial handshake, accompanied by an encrypted pseudo-identity $CID_{D_i}^{T_2}$. Upon decryption, $CID_{D_i}^{T_2}$ reveals the $PID_{D_i}$, keeping the original identity $OID_{D_i}$ hidden. Furthermore, $PID_{D_i}$ is refreshed in each session, making it substantially harder for adversaries to deduce or exploit the node’s true identity.
\subsubsection{Traceability and revocation}
For any two communicating entities, $RCA$ and $D_i$, only $RCA$ can link $PID_{D_i}$ to its corresponding certificate $Cert_{D_i}$ from the list of communicating devices $list_{Comm}$. Thus, $RCA$ can report $D_i$'s misbehaviour to the $CA$ by submitting its certificate. Since only the $CA$ can associate $Cert_{D_i}$ with the $OID_{D_i}$, the node can be easily traced and revoked by adding its certificate to the CRL, demonstrating that the proposed scheme ensures traceability and revocation.
\subsubsection{Unlinkability}
The proposed scheme enhances privacy by using session-specific $PID_{D_i}^{T_i}$. Each device $D_i$ responds to the $RCA$ with $PID_{D_i}$ instead of its original identity $OID_{D_i}$, which is encrypted using $SK_{RCA-D_i}$. Since $PID_{D_i}$ is dynamically updated every session period $T_s$, it becomes highly difficult for an adversary to correlate messages across different sessions, achieving unlinkability between different transmitted messages every $T_s$.
\subsubsection{Non-Repudiation}
Digital signatures are vital for ensuring message authenticity. Each entity holds an ECC-based public/private key pair. During initial authentication, messages are signed with the sender’s private key, ensuring origin authentication and integrity, while preventing denial of transmission due to the cryptographic proof. During reauthentication, the pseudo-identity $PID_{D_i}^{T_3}$ is encrypted as $CID_{D_i}^{T_3} = ENC_{SK_{RCA-D_i}}(PID_{D_i}^{T_3})$, using a symmetric key $SK_{RCA-D_i}$ derived via Diffie-Hellman exchange: $SK_{RCA-D_i} = SK_{RCA} \cdot PK_{D_i}$. Since only the entity knows its private key, unauthorized access or forgery is prevented. Additionally, each device is uniquely characterized by its CFO and quadrature skewness signature—non-forgeable PHY-layer features tied to hardware imperfections—ensuring strong non-repudiation during both registration and reauthentication.
\subsubsection{Security robustness against active attacks}
This part demonstrates the security robustness of the proposed scheme as follows.
\begin{enumerate}
    \item \textbf{Robustness against replay attack}: The proposed scheme effectively resists replay attacks by combining timestamp verification with non-forgeable physical-layer features. Each transmitted message includes a timestamp, and the receiver checks its freshness using $T_r - T_i \leq \Delta T$; outdated or duplicate timestamps lead to immediate rejection. Additionally, during reauthentication, PHY-layer features such as CFO and skew, extracted from received OFDM symbols, act as unique hardware fingerprints. These cannot be replicated through simple replay. Messages replayed from different devices are detected and rejected by the device verification model, ensuring a strong defense at both cryptographic and hardware levels against replay attacks.

\item \textbf{Robustness against modification attack}: The proposed scheme ensures strong resilience against MitM and message modification attacks through layered cryptographic and physical defenses. Each entity ($D_i$, $RCA$, and $CA$) possesses a digital certificate signed by the $CA$. These certificates are verified during the initial handshake, and any alteration invalidates the signature, preventing tampering. Session keys are securely established using ECDH key exchanging protocols as $SK_{RCA-D_i} = SK_{D_i} \cdot PK_{RCA} = SK_{RCA} \cdot PK_{D_i}$, ensuring that intercepted messages cannot be decrypted or modified without the private keys. During reauthentication, robustness is further enhanced via PHY-layer fingerprinting. The $RCA$ extracts device-specific CFO and skew features from received OFDM symbols using the Van-de-Beek algorithm. As these features are hardware-dependent and non-replicable, a MitM attacker cannot mimic them, making spoofing and modification infeasible and ensuring a strong second layer of defense.
\item \textbf{Robustness against impersonation attack}:
The proposed scheme resists impersonation attacks by integrating cryptographic verification with PHY-layer device fingerprinting. During initial authentication, each entity—including $D_i$ and $RCA$—receives a long-term digital certificate signed by the $CA$, binding public keys to their owners. These certificates, verified alongside digital signatures and timestamps during the handshake, prevent unauthorized access and make it infeasible to reuse or fabricate valid credentials. In the reauthentication phase, robustness is reinforced by hardware-level verification. CFO and skew values, extracted from OFDM symbols using the Van-de-Beek algorithm, serve as unique device fingerprints. These features are input to a machine learning classifier for identity verification and cannot be forged without access to the physical hardware, rendering impersonation ineffective even if cryptographic keys are compromised. Additionally, session-specific pseudo-identities, encrypted with symmetric keys, are periodically refreshed. This prevents session correlation and tracking, enhancing anonymity and further mitigating impersonation threats.

\begin{figure*}[t!]
\centerline{\includegraphics[width=17cm]{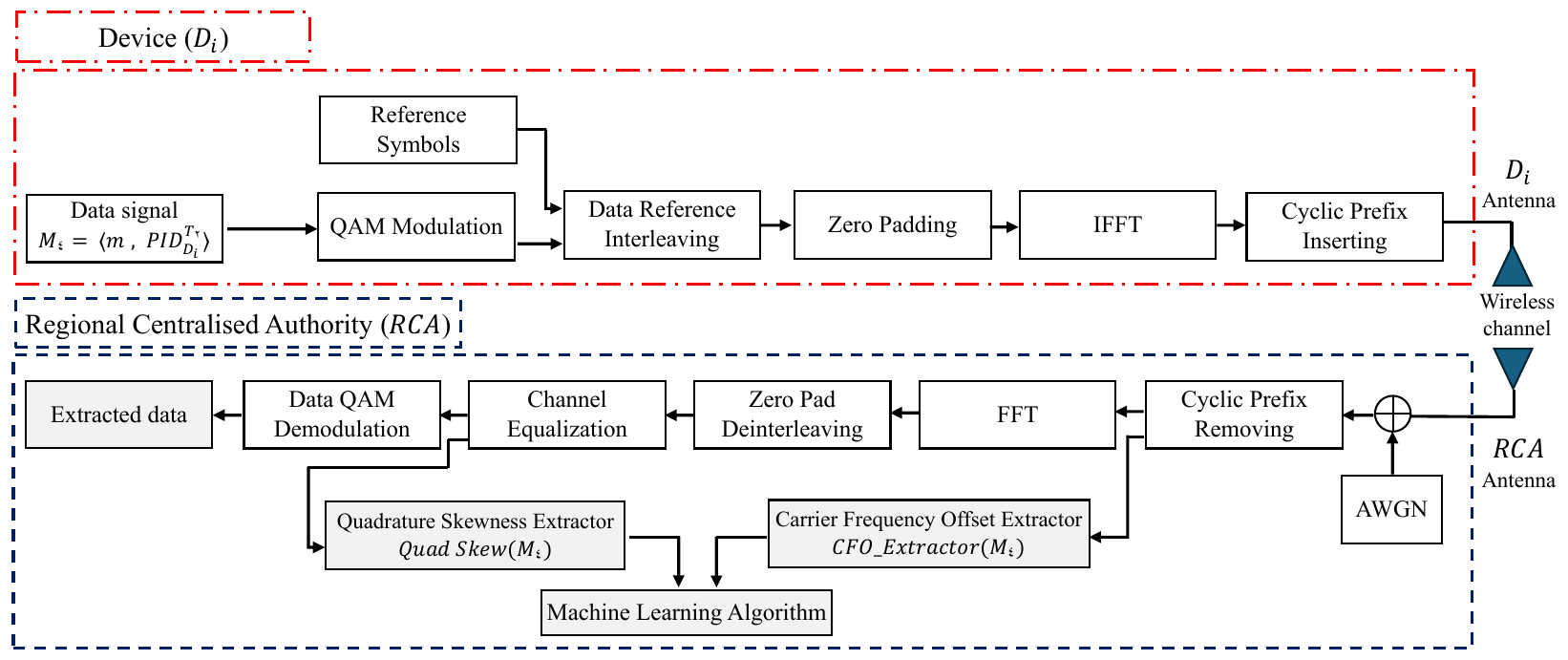}}
\setlength\abovecaptionskip{0.5\baselineskip}
\caption{Block diagram of the neural network-enhanced OFDM system simulation for indoor 5G by labview}
\label{f_lab_view}
\end{figure*}

\end{enumerate}

\subsection{Evaluation of PHY-Layer Mechanisms}
This part analyzes the performance metrics associated with the PHY-layer re-authentication procedure, as follows.
\subsubsection{Simulation analysis}\label{S4.1}
This comprehensive simulation evaluates the performance of an OFDM system tailored for indoor 5G environments, with a particular focus on the estimation of $D_i$'s CFO and quadrature skewness using \textit{CFO\_Extractor} and \textit{Quad\_Skew} functions. The system is configured with $256$ subcarriers, a $5\ MHz$ IQ sampling rate, and 4-QAM modulation to ensure robust data transmission under realistic operational conditions. To emulate practical indoor propagation, the channel model incorporates $10$ multipaths with Rayleigh fading channel with a uniformly distributed phase response $U[-\pi, \pi)$, and additive white Gaussian noise (AWGN) added at the receiver side. This channel model is modelled between two communicating ebtities a transmitter $Tx$ represents $D_i$ and a receiver represents $RCA$. The channel is modeled between two communicating entities: a transmitter ${Tx}$, representing the node $D_i$, and a receiver $Rx$ representing the RCA. The SNR values is set to $5$ dB and IQ gain imbalance and sample offset are intentionally excluded to isolate the effects of CFO and skewness. Accordingly, two scenarios are examined using the previously specified configuration settings to generate a dataset for estimating CFO and quadrature skewness (in degrees) for a designated number of simulated devices. The scenarios are defined as follows:
\begin{enumerate}
    \item \textit{CFO Estimation}: The simulation employs settings with a constant quadrature skewness of 3 degrees.
    \item \textit{Quadrature Skewness Estimation}: The simulation utilizes settings with a fixed CFO of 50 normalized units.
\end{enumerate}
These controlled scenarios facilitate a precise assessment of the machine learning classifier’s capability to differentiate between these specific impairments.
The simulation framework, illustrated in Fig. \ref{f_lab_view}, consists of three core components: the transmitter, channel, and receiver. At the transmitter, random binary data is generated, modulated using 4-QAM, and transformed from serial-to-parallel format for inverse fast Fourier transform (IFFT) processing. The IFFT converts frequency-domain symbols into time-domain signals, followed by the addition of a CP to mitigate inter-symbol interference. CFO and quadrature skewness impairments are then introduced, either as fixed or variable parameters, depending on the scenario.

In the channel model, the signal undergoes multipath propagation and Rayleigh fading, with random phase shifts applied and noise added to simulate realistic indoor conditions. At the receiver, the cyclic prefix is removed, the time-domain signal is transformed back to the frequency domain via FFT, and the data is converted back to serial format before demodulation.

\subsubsection{ML evaluation for optimal model selection}
\label{S4.2.3}

\begin{figure}[t!]
\centerline{\includegraphics[width=8.5cm]{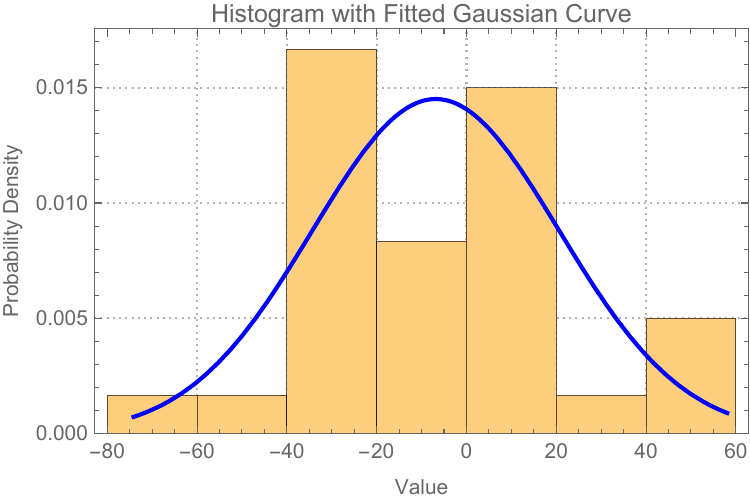}}
\centerline{(a) CFO histogram}
\centerline{\includegraphics[width=8.5cm]{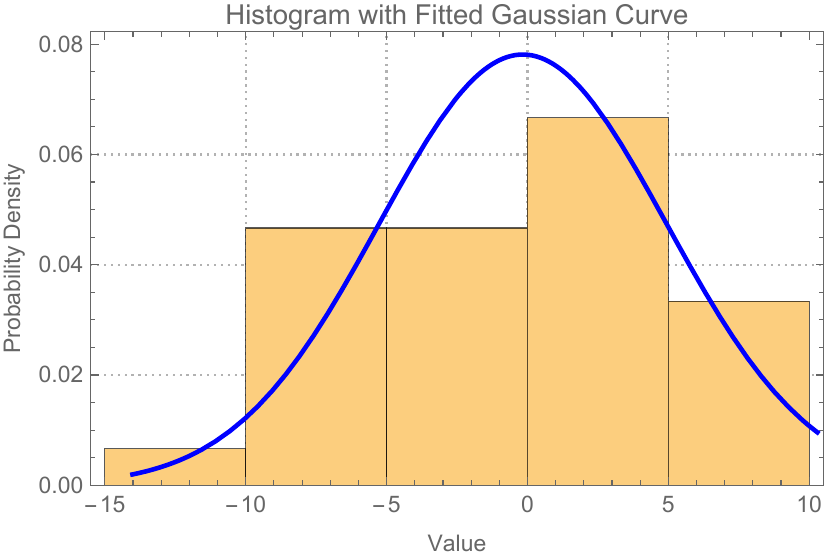}}
\centerline{(b) Quad skewness histogram}
\setlength\abovecaptionskip{0.5\baselineskip}
\caption{Histograms with fitted gaussian curve}
\label{f442}
\end{figure}

\begin{table*}
   \caption{ML results}
   \centering
   \resizebox{.95\textwidth}{!}{
    \begin{tabular}{c|c|c|c|c|c|c|c|c|c}
        \hline
       \multirow{2}{*}{\textbf{ML Algorithm}} & \multicolumn{3}{c|}{\textbf{Accuracy}} & \multicolumn{3}{c|}{\textbf{Recall Mean}} & \multicolumn{3}{c}{\textbf{F1 Score Mean}} \\
        \cline{2-10}
        & 10 Devices & 20 Devices & 30 Devices & 10 Devices & 20 Devices & 30 Devices & 10 Devices & 20 Devices & 30 Devices \\
        \hline
        Logistic Regression & 86.3\% & 87.95\% & 74.75\% & 0.863 & 0.8795 & 0.7473 & 0.8612 & 0.8786 & 0.7473 \\
        Neural Network & 89.25\% & 88.975\% & 77.5667\% & 0.8925 & 0.8897 & 0.7756 & 0.8895 & 0.8888 & 0.7759 \\
        Decision Tree & 89.45\% & 88.6\% & 77.65\% & 0.8945 & 0.886 & 0.7765 & 0.8922 & 0.8854 & 0.7760 \\
        Gradient Boosted Trees & 89.4\% & 89.1\% & 77.85\% & 0.896 & 0.891 & 0.7793 & 0.8952 & 0.8904 & 0.7815 \\
        Nearest Neighbor & 89.25\% & 89.85\% & 78.75\% & 0.8925 & 0.8985 & 0.7875 & 0.8914 & 0.8978 & 0.7875 \\
        \hline
    \end{tabular}}
    \label{tab:my_label}
\end{table*}
% Introducing the evaluation of machine learning algorithms
This study assesses the performance of five machine learning algorithms—logistic regression, neural networks, decision trees, gradient boosted trees, and K-nearest neighbors (KNN)—in estimating impairments within an OFDM system designed for indoor 5G environments. The evaluation was conducted under standardized conditions, with simulations involving $10$, $20$, and $30$ devices. Key performance metrics—accuracy, mean recall, and mean F1 score—were employed to evaluate the algorithms’ ability to accurately classify impairments, such as CFO and quadrature phase imbalance (skewness), under challenging conditions including multipath propagation, Rayleigh fading, and low SNR, as detailed in Section \ref{S4.1}. The distributions of these impairments for $30$ devices are illustrated in the histograms provided in Fig. \ref{f442} (a) and (b) for CFO and quadrature skewness, respectively. The CFO histogram reveals a distribution with a mean of $-6.883$ and a variance of $658,450$, indicating a slight negative bias and moderate variability across devices. Meanwhile, the quadrature skewness histogram exhibits a negatively skewed distribution with a mean of $-6.883$ and a variance of $758,450$, reflecting significant spread and a tendency toward more extreme negative values. The results, summarized in Table \ref{tab:my_label}, provide insights into the scalability and robustness of each algorithm across varying device density scenarios.

Logistic regression, a linear classification model, predicts the probability of a binary outcome and can be extended to multiclass classification tasks. Its computational efficiency and interpretability make it suitable for simpler datasets, though its performance may degrade with complex, non-linear relationships \cite{M1}. For 10 devices, logistic regression achieves an accuracy of 86.3\%, a mean recall of 0.863, and a mean F1 score of 0.861297. With 20 devices, these metrics improve slightly to 87.95\%, 0.8795, and 0.8786, respectively. However, with 30 devices, performance declines significantly, yielding an accuracy of 74.75\%, a mean recall of 0.7473, and a mean F1 score of 0.7473.

Neural networks, inspired by biological neural systems, excel at modeling complex, non-linear patterns but are computationally intensive and require substantial training data \cite{M2}. For 10 devices, neural networks attain an accuracy of 89.25\%, a mean recall of 0.8925, and a mean F1 score of 0.8895. With 20 devices, the metrics are 88.975\%, 0.8897, and 0.8888, respectively. For 30 devices, performance drops to an accuracy of 77.5667\%, a mean recall of 0.7756, and a mean F1 score of 0.7759.

Decision trees, a non-parametric method, utilize a tree-like structure to make decisions based on feature values, offering intuitiveness and the ability to handle both numerical and categorical data. However, they are prone to overfitting without proper pruning \cite{M1}. For 10 devices, decision trees achieve an accuracy of 89.45\%, a mean recall of 0.8945, and a mean F1 score of 0.8922. With 20 devices, these metrics are 88.6\%, 0.886, and 0.8854, respectively. For 30 devices, performance decreases to an accuracy of 77.65\%, a mean recall of 0.7765, and a mean F1 score of 0.7760.

Gradient boosted trees, an ensemble technique, combine multiple decision trees, with each tree correcting the errors of its predecessors, making them highly effective for complex datasets but computationally demanding \cite{M3}. For 10 devices, gradient boosted trees record an accuracy of 89.4\%, a mean recall of 0.896, and a mean F1 score of 0.8952. With 20 devices, the metrics are 89.1\%, 0.891, and 0.8904, respectively. For 30 devices, performance declines to an accuracy of 77.85\%, a mean recall of 0.7793, and a mean F1 score of 0.7815.

K-nearest neighbors, a non-parametric algorithm, classifies instances based on the majority class of their k nearest neighbors in the feature space, offering adaptability to local data structures without distributional assumptions \cite{M1}. For 10 devices, KNN achieves an accuracy of 89.25\%, a mean recall of 0.8925, and a mean F1 score of 0.8914. With 20 devices, these metrics improve to 89.85\%, 0.8985, and 0.8978, respectively. For 30 devices, KNN outperforms other algorithms with an accuracy of 78.75\%, a mean recall of 0.7875, and a mean F1 score of 0.787597.

\begin{figure}[t!]
\centerline{\includegraphics[width=8.5cm]{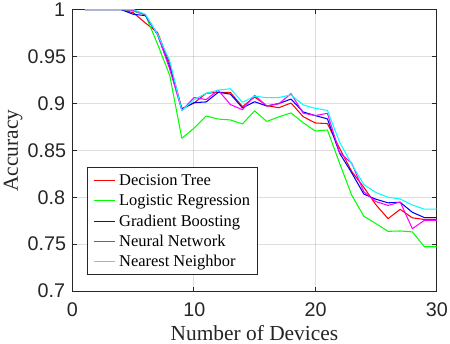}}
\setlength\abovecaptionskip{0.5\baselineskip}
\caption{Accuracy across different number of devices}
\label{f44}
\end{figure}

As presented in Table \ref{tab:my_label}, all algorithms demonstrate robust performance for 10 and 20 devices, with accuracies ranging from 86.3\% to 89.85\%. However, with 30 devices, a notable decline is observed, with accuracies falling to between 74.75\% and 78.75\%. This performance degradation is likely due to increased complexity, such as heightened interference or data variability, challenging the algorithms’ generalization capabilities. Among the evaluated algorithms, KNN achieves the highest accuracy (78.75\%), mean recall (0.7875), and mean F1 score (0.7875) for 30 devices, surpassing gradient boosted trees (77.85\%, 0.7793, 0.7815), decision trees (77.65\%, 0.7765, 0.7760), neural networks (77.5667\%, 0.7756, 0.7759), and logistic regression (74.75\%, 0.7473, 0.7473), representing (accuracy, recall, F1 score), see Fig. \ref{f44}. This superior performance underscores KNN’s ability to effectively identify positive instances and balance precision and recall in high-density scenarios.

\begin{figure*}[t!]
\centerline{\includegraphics{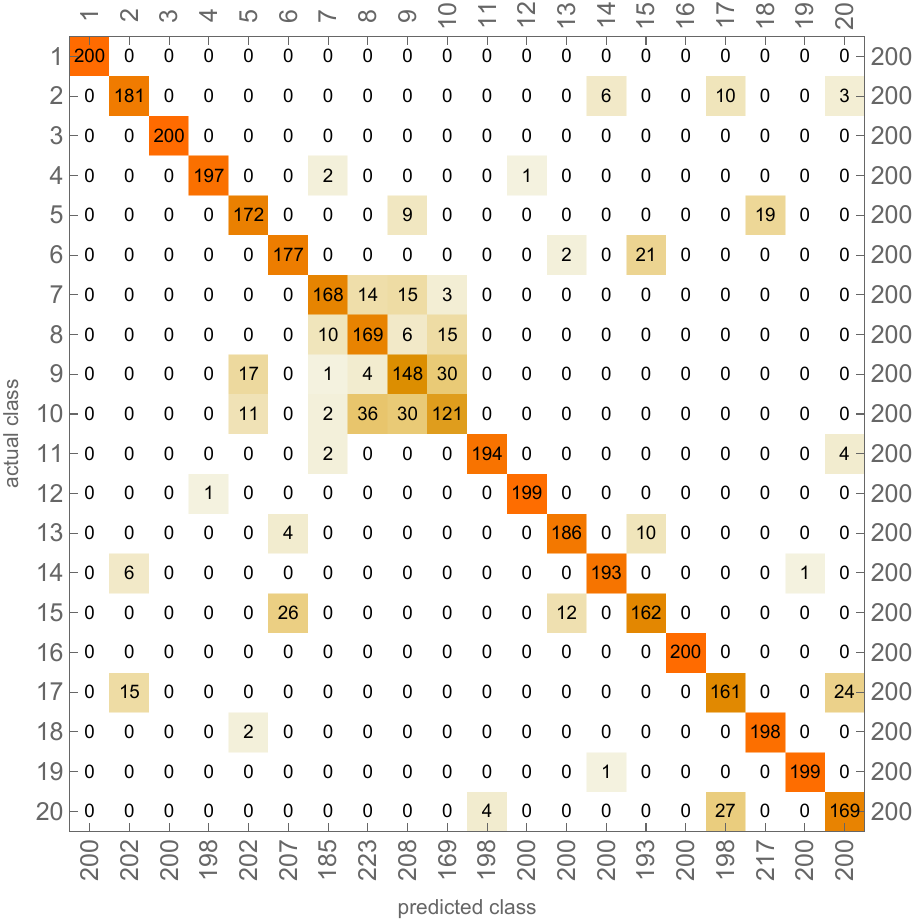}}
\setlength\abovecaptionskip{0.5\baselineskip}
\caption{Confusion matrix of nearest neighbor algorithm for 20 devices scenario}
\label{f3}
\end{figure*}

The selection of KNN for this application is driven by its high accuracy and robust performance with 30 devices. Unlike parametric models such as logistic regression, which struggle with non-linear relationships, or complex models like neural networks, which may overfit or demand extensive computational resources, KNN adapts to the local data structure without distributional assumptions. This adaptability is particularly advantageous in indoor 5G environments, where increased device counts introduce greater variability and interference. Additionally, KNN’s computational efficiency supports its practicality for real-time impairment estimation in OFDM systems, as corroborated by \cite{M4}, which emphasizes the need for low-complexity solutions in 5G small cells. Fig. \ref{f3} presents the confusion matrix for the 20-device scenario using KNN, further illustrating its performance.

It is observed that while all algorithms exhibit competitive performance for smaller device counts, KNN emerges as the optimal choice for estimating impairments in the proposed indoor 5G OFDM system due to its superior accuracy and robustness with 30 devices. Its ability to handle increased complexity makes it well-suited for practical deployment in high-density 5G environments.

\subsection{Computation and Communication Overheads }\label{S5.2}
In the context of $HIE$ within intra-hospital networks, the proposed cross-layer authentication scheme integrates cryptographic operations, PHY-layer security features, adaptive security policies, and periodic pseudo-identity refreshes to ensure robust security while maintaining computational efficiency—a critical requirement for deployment in resource-constrained medical devices and real-time healthcare applications. To evaluate the scheme’s practicality, we conduct a detailed analysis of computation and communication overheads, benchmarking against state-of-the-art approaches, the analysis draws on typical overheads for similar schemes and qualitative comparisons with existing protocols to provide a comprehensive evaluation. The comparison of computation and communication overheads is done between our work , Qi-Xie et al. \cite{b22}, Xiang et al. \cite{b29}, Kumar et al. \cite{b30} and Chen et al. \cite{b31}.
  
\subsubsection{Computation overhead}\label{S5.2.1}
Computation overhead refers to the processing time required to execute authentication tasks, measured in \textit{(ms)} on typical healthcare $IoT$ devices. The proposed scheme involves several computational components, each contributing to the overall overhead. Table \ref{excu} shows the execution time for each cryptographic operation according to \cite{b32}, the simulation of computational cost is done using MIRCAL library \cite{b33} under the following condition the desktop is quad-core i7-4790 CPU, 16 GB RAM, and the operating system is Linux Ubuntu 20.04-desktop-amd64, and Raspberry PI 3B is quad-core ARM Cortex-A53, 1 GB RAM.
\begin{table}[h]
\caption{Notation for execution time parameter in ($ms$) \cite{b32}}
\centering\label{excu}
\renewcommand{\arraystretch}{1.3}
\resizebox{0.48\textwidth}{!}{
\begin{tabular}{c|c|c} 
\hline
\textbf{Symbol} & \textbf{Description} & \textbf{Time ($ms$)} \\
\hline
$T_{ecc}^{mul}$ & Time for $ECC$-based scalar multiplication & 1.489 \\
$T_{ecc}^{add}$ & Time for $ECC$-based group addition & 0.008 \\
$T_{h}$ & Time for hashing & 0.003 \\
$T_{Enc}$ & Time for encryption ($AES-256$) & 0.002 \\
$T_{Dec}$ & Time for decryption ($AES-256$) & 0.001 \\
\hline
\end{tabular}}
\end{table}

\begin{figure}[htp]
\centerline{\includegraphics[width=8cm]{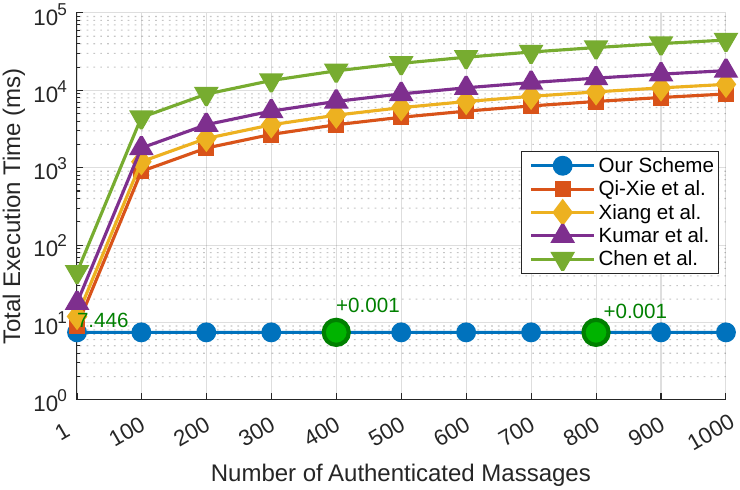}}
\setlength\abovecaptionskip{0.5\baselineskip}
\caption{\centering{Comparison of computation overhead with related schemes}}
\label{f99}
\end{figure}

\begin{table*}[t]
    \centering
    \caption{Computation of overhead and communication cost for $n$ Authenticated messages}\label{tab:comparison}
   \resizebox{.9\textwidth}{!}{
    \begin{tabular}{c|c|c}
       \hline
       \textbf{Scheme} & \textbf{Authentication Overhaed} &\textbf{Authentication Cost} \\

      \hline
Qi-Xie et al. \cite{b22}&$13nT_{h}+6nT_{ecc}^{mul}+2nT_{Enc}+2T_{Dec} =8.979n \ ms$&$661n \ bytes$ \\
Xiang et al. \cite{b29}&$12nT_{h}+8nT_{ecc}^{mul}=11.948n \ ms$&$240n \ bytes$\\
Kumar et al. \cite{b30}&$26nT_{h}+12nT_{ecc}^{mul}+4nT_{Enc}+4nT_{Dec} =17.985 \ ms$&$984n \ bytes$\\
Chen et al. \cite{b31}&$ 30nT_{ecc}^{mul}+2nT_{Enc}+2nT_{dec}=44.676n \ ms$&$400n \ bytes$\\
     Our's & $5T_{ecc}^{mul}+\left\lceil \frac{n}{d} \right\rceil T_{dec} =7.445+0.001\left\lceil \frac{n}{d} \right\rceil \ ms$ &\(144+20n+256\left\lceil \frac{n}{d} \right\rceil \ bytes \)\\   
         \hline
    \end{tabular}}
    
\end{table*}
For the proposed scheme, the execution time for the initial handshaking process involves the certificate verification ($2\times$ multiplication on ECC), signature verification ($2\times$ multiplication on ECC), Diffi-Helman exchange for secret key (single multiplication on ECC) and a single symmetric decryption. Thus, the total processing time of the verification process in the initial handshaking is $ 5T_{ecc}^{mul}+T_{dec} \equiv(5\times1.489+0.001=7.446 \ ms)$. For re-authentication, the overhead is decreased, saving much more of the overheads associated with the other four schemes, due to re-authentication phase of our scheme which is dependent on unique features of each device the CFO and quadrature skewness extraction and applied to trained algorithm $ML$ (KNN), the overall overhead cost of this process will be fraction of $\mu sec$ which is negligible against other ECC cryptographic operation. In the re-authentication process, we update the pseudo identity every session time where the $RCA$ receives the tuple  $M_5 = \langle m, {PID}_{D_i}^{T_2}, {CID}_{D_i}^{T_3} \rangle$  which requires only decryption operation and that's for every session time $T_s$, if the received messages during $T_s$ is $d$ messages. Then, the computational overhead for the re-authentication of ($n$) massages is \(\left\lceil \frac{n}{d} \right\rceil\times T_{Dec} = ( \left\lceil \frac{n}{d} \right\rceil\times0.001 ) \ ms \) added to the initial authentication overhead, thus the total overhead $=7.445+0.001\left\lceil \frac{n}{d} \right\rceil \ ms$.
for Qi-Xie et al. \cite{b22} the overhead of $n$ authentication denoted as $13nT_{h}+6nT_{ecc}^{mul}+2nT_{Enc}+2nT_{Dec} \equiv(13n\times0.003+6n\times1.489+2n\times0.002+2n\times0.001=8.979n \ ms)$, for Xiang et al. \cite{b29} specific protocol the overhead is $12nT_{h}+8nT_{ecc}^{mul}\equiv(12n\times0.003+8n\times1.489=11.948n \ ms)$, Kumar et al. \cite{b30} protocol patients and doctors hold each other’s public keys and negotiate a session key through the cloud server, so that's involves in multiple symmetric encryptions, $ECC$ point multiplications and signatures, single authentication overhead is $26nT_{h}+12nT_{ecc}^{mul}+4nT_{Enc}+4nT_{Dec} \equiv(26n\times0.003+12n\times1.489+4n\times0.002+4n\times0.001=17.985n \ ms)$. Finally for Chen et al. \cite{b31} scheme, users establish a session key with a sensor node via the gateway. The scheme's security relies on hash functions and pre-shared secrets and it does not provide forward security single authentication overheads around $ 30nT_{ecc}^{mul}+2nT_{Enc}+2nT_{dec} \equiv(30n\times1.489+2n\times0.002+2n\times0.001=44.676n \ ms)$. As calculated before our scheme shows the least computational overhead against the other four scheme we compare against. Fig. \ref{f99}, illustrates the total computation cost analysis for $1000$ authenticated massages between our scheme and others schemes.

\begin{figure}[htp]
\centerline{\includegraphics[width=8cm]{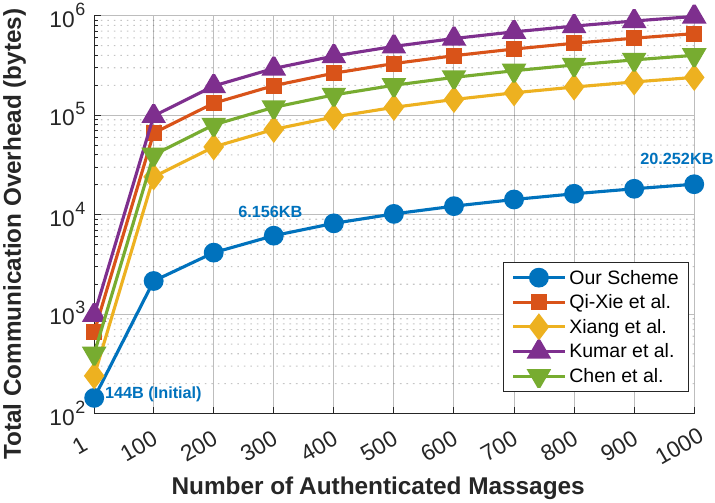}}
\setlength\abovecaptionskip{0.5\baselineskip}
\caption{\centering{Comparison of communication overhead with related schemes}}
\label{f101}
\end{figure}

\subsubsection{Communication overhead}
\label{S5.2.2}

Communication cost refers to the amount of data exchanged between entities during the execution of a protocol or operation. It is typically measured in terms of the number of transmitted messages and the size of each message, by analyzing the communication cost of the proposed scheme, comparing it to related approaches in terms of message transmitted size. Since our scheme is a cross-layer authentication scheme, reducing communication cost during the reauthentication phase. The curve type ``$SS512$'' \cite{b34} is selected for the security level of 80 bits. Table \ref{tab:my_recom} describes the element size in the group ECC based on $G$ which is 320 bits (40 bytes), for the element size in $Z_q$ is 160 bits (20 bytes), the cost of the timestamp is 32 bits (4 bytes), and finally the AES-256 is 256 bits \cite{b35}, the cost of the message is ignored as the massage is constant for all schemes, Table \ref{tab:my_bitlength}, defines the bit of the proposed scheme. The received tuple at the initial authentication for verification is $M_3 = \langle Cert_{D_i}, T_2, CID_{D_i}^{T_2}, \sigma_{D_i} \rangle$ \(= \langle \{PK_{D_i}, \ T_R, \ \sigma_{CA}\}, \ T_2, \ CID_{D_i}^{T_2}, \ \sigma_{D_i}\rangle \), as \(PK_{D_i}\in G \) so it costs 40 bytes, for \( \{\sigma_{CA}, \ CID_{D_i}^{T_2}, \ \sigma_{D_i} \} \) each cost 32 bytes, and for the timestamp 4 bytes for each, therefore, the total cost is \(40+3\times32+ 2\times 4=144 \ bytes \ (1152 \ bit)\). The re-autentication uses two functions only for verification $CFO_{Extractor}$ and $Quad_{Skweness}$ at the received tuple is $M_4 = \langle m, {PID}_{D_i}^{T_2} \rangle $ which its communication overhead only the ${PID}_{D_i}^{T_2}$ cost, also the pseudo identity need to be updated every $T_s$ by receiving tuple $M_5 = \langle m, {PID}_{D_i}^{T_2}, {CID}_{D_i}^{T_3} \rangle$. To calculate the overall communication cost for $n$ messages, as assumed before the number of received messages during $T_s$ is $d$. So that the total communication cost equal to\((144+ 20\times n + \left\lceil \frac{n}{d} \right\rceil\times256)=(144+20n+256\left\lceil \frac{n}{d} \right\rceil) \ bytes \). For Qi-Xie et al. \cite{b22}, the verification phase has a total communication overhead of around 5288$n$ bits ($661n\ bytes$). The Xiang et al. \cite{b29}, consumes 1920$n$ bits ($240n\ bytes$) for verification. In Kumar et al. \cite{b30}, upon verification, the server delivers an enormous amount of encrypted data, leading to an overall overhead of around 7872$n$ bits($984n\ bytes$). For Chen et al. \cite{b31} scheme, the overall overhead in verification is 3200$n$ bits ($400n\ bytes$). Table \ref{tab:comparison}, shows the comparison between the proposed scheme and other's scheme for $n$ authenticated massages. Fig. \ref{f101}, illustrates the total communication cost analysis for $1000$ authenticated massages between our scheme and others schemes.

\begin{table}[t]
    \centering
   \caption{Bit length of each element in our experiment}
    \resizebox{.48\textwidth}{!}{\label{tab:my_bitlength}
    \begin{tabular}{c|c|c}
    \hline
       \textbf{Parameter}  & \textbf{Description} & \textbf{Size (bits)} \\
       \hline
       $G$  & Bit length of an element in $G$ & 320\\
       $Z_q$& Bit length of an element in $Z_q$ & 160\\
      $PID$& Bit length of pseudo identity & 160\\
       $T$& Bit length of a timestamp & 32\\
       $CID$& Bit length of the AES cipher text & 256\\
       \hline
    \end{tabular}}
    
\end{table}

\section{Conclusion}\label{S7}
This paper presented a cross-layer authentication scheme tailored for enhancing security within health information exchange networks. By integrating ECC-based cryptographic techniques with PHY-layer feature extraction and machine learning-based re-authentication, the proposed system delivers continuous, lightweight identity verification while preserving low computational and communication overheads. The scheme addresses critical security threats such as impersonation, MitM, replay, and modification attacks, and demonstrates robustness through formal validation using BAN logic. Simulation results confirms the practicality of the proposed approach, particularly in high-density, resource-constrained medical environments, where reliable and efficient authentication is imperative. In addition to its strong security guarantees, the adoption of frequently refreshed encrypted pseudo-identities ensures enhanced privacy and resistance to identity tracking. Future work will focus on expanding the model to support decentralized trust management through blockchain integration, enabling further scalability and transparency. Additionally, extending the PHY-layer feature set using deep learning-based representation learning and incorporating federated learning techniques may enhance accuracy and privacy in large-scale, multi-institutional deployments. Real-world deployment and clinical validation will also be pursued to assess operational performance in diverse healthcare settings.

\bibliographystyle{ieeetran}
\bibliography{biblio_traps_dynamics}

\end{document}